# A mutate-and-map protocol for inferring base pairs in structured RNA

Pablo Cordero[1,‡], Wipapat Kladwang[2,‡], Christopher C. VanLang[3], and Rhiju Das[1,2,4*]

*Departments of Biomedical Informatics[1], Biochemistry[2], Chemical Engineering[3], and Physics[4], Stanford University, Stanford CA 94305*

* To whom correspondence should be addressed: rhiju@stanford.edu. Phone: (650) 723-5976. Fax: (650) 723-6783.

ꝉ Contributed equally

## *Summary*

Chemical mapping is a widespread technique for structural analysis of nucleic acids in which a molecule's reactivity to different probes is quantified at single-nucleotide resolution and used to constrain structural modeling. This experimental framework has been extensively revisited in the past decade with new strategies for high-throughput read-outs, chemical modification, and rapid data analysis. Recently, we have coupled the technique to high-throughput mutagenesis. Point mutations of a base-paired nucleotide can lead to exposure of not only that nucleotide but also its interaction partner. Carrying out the mutation and mapping for the entire system gives an experimental approximation of the molecules "contact map." Here, we give our in-house protocol for this "mutate-and-map" strategy, based on 96-well capillary electrophoresis, and we provide practical tips on interpreting the data to infer nucleic acid structure.

## *Introduction*

Elucidating the structures of nucleic acids is essential to understanding the mechanisms that govern their many biological roles, ranging from packing and transmission of genetic information to regulation of gene expression [e.g. refs. (*1–3*)]. Many biophysical, biochemical, and phylogenetic methods have been developed to probe nucleic acid folds (*4–6*), but rapid determination of a functional RNA's secondary and tertiary structures remains generally challenging. Amongst available methods, chemical and enzymatic mapping (or "footprinting") approaches are particularly facile. First, the molecule is covalently modified by the mapping reagent at nucleotides that are exposed or unstructured. The modifications can then be detected in various ways, including reverse transcription: here, the adducts formed by modification will stop the reverse transcriptase, generating complementary DNA (cDNA) of specific lengths that highlight the molecule's unstructured sites and that can be read out by gel or capillary electrophoresis. Chemical mapping can provide a range of useful information: DNA and RNA can both be interrogated along their Watson-Crick base edges through alkylation or formation of other adducts (*7*, *8*). Additionally, RNA can be probed through 2´ hydroxyl acylation or metal-ion-catalyzed 2´ hydroxyl in-line cleavage at phosphates to reveal "dynamic" nucleotides (*9*, *10*), or to determine accessibility of the phosphate (*11*) or more complex features such as participation in helices (*12–15*). DNA/RNA duplexes or DNA structures can be similarly analyzed (5), although this chapter will focus on RNA cases where the method has been most thoroughly tested.

Chemical mapping is applicable to essentially any RNA, including large nucleic acid systems *in vivo* or *in vitro* (*16–18*). However, while a useful source of information, the one-dimensional nature of the resulting data (one reactivity value per nucleotide) limits the conclusions that can be deduced from these experiments, as several structural hypothesis can be consistent with the detected reactivity data (*19–21*). To help address these ambiguities, we developed a "mutate-and-map" strategy whereby each nucleotide of the RNA is systematically mutated and its interaction partners are discovered by their "release" and increase in chemical accessibility (*22–24*). Chemical mapping of single-point mutants reveal the molecule's approximate "contact map" and can be quickly performed using current molecular-biology tools. Automated analysis protocols can then quantitate these data and inform structural models. The data appear particularly constraining at the level of secondary structure. Furthermore, in several cases, unambiguous information related to tertiary contacts or pseudoknots formed by the RNA are obtained. For some cases, the single mutations produce substantial perturbations to the structure, but these are straightforward to detect as perturbations in the chemical reactivity profiles, and, indeed, reflect the propensity of the wild type RNA to switch its structure, a property of functional interest. Overall, the mutate-and-map method provide rich data sets with information content that generally exceeds prior chemical or enzymatic approaches to RNA structure.

The mutate-and-map experimental protocol begins by assembling DNA templates corresponding to the RNA mutants of interest, using PCR assembly (*23*). The DNA is then purified and *in vitro* transcribed. The resulting RNA is folded or, for DNA/RNA or other multicomponent systems, assembled into the desired complex, and then subjected to standard chemical mapping protocols using fluorescently labeled primers and capillary electrophoresis. Each purification step is carried out with magnetic beads and multichannel pipetters in 96-well format, enabling the probing of hundreds of mutants in parallel. The current requirement for synthesis via PCR assembly limits the length range of the current procedure to approximately 300 nucleotides. However, after a correct PCR assembly is found, the complete protocol can be performed on the timescale of two days: each major step in the procedure (PCR assembly and DNA purification; transcription and RNA purification; chemical mapping and reverse transcription; and capillary electrophoresis) takes on the order of 3 to 5 hours. Quantitation of the data and its use in secondary structure inference is similarly rapid. More sophisticated analyses of the data as well as accelerations from next-generation synthesis and sequencing methods have been piloted in our laboratory, but the focus of this chapter will be on our stable in-house protocol which has been applied to dozens of RNA sequences at the time of writing.

## *Materials*

### **General materials and equipment**

All the procedures described are performed in a 96-well plate format. Each sample plate contains 96 mutants; so if the sequence to be probed has 119 nucleotides, then the wild type construct plus all single-nucleotide mutants will occupy 120 wells, or one and a quarter sample plates. Multichannel pipetters are used to reduce the pipetting effort required in each step.

1. At least 15 boxes of extra-long P10 tips per sample plate; 3-4 boxes of P100 refill tips (E&K Scientific) per plate; at least one box each of P20 filter tips, and P200 filter tips for premixes.

2. Several 96-well V-bottom plates (Greiner); at least 4 per sample plate. In many steps, an "auxiliary" plate will be used to pre-aliquot premixes and reduce pipetting effort.

3. Plastic plate seal films (EdgeBio), at least three per sample plate used.

4. 100 mL 70% ethanol per sample plate (see Note 1)

5. 96 post magnetic stands, one per sample plate (available at VP Scientific)

6. 0.5 mL dNTP mixture [10 mM each dATP, dCTP, dTTP, dGTP, prepared from 100 mM stocks of each dNTP (New England Biosystems)] per sample plate.

7. 2 mL of Agencourt AMPure XP beads (Beckman Coulter) for nucleic acid purification.

8. Two 96-well format dry incubators (Hybex, SciGene) set to 42 °C and 90 °C before starting the chemical mapping part of the experiment. If these types of incubators are not available, a water bath can be used.

9. 50 mL RNAse-free sterile water. We use Barnstead MilliQ purified water.

10. Access to an ABI capillary electrophoresis sequencer. We use an ABI3100 sequencer (16 capillary) for local tests and send samples to an external company (Elim Biosciences, Hayward, CA) for higher throughput runs on 96-capillary ABI 3730 sequencers.

## Sequence assembly and fluorescent primers

The mutate-and-map method requires a fluorescent primer that will be used to reverse transcribe the RNA to detect modification sites and a working set of DNA primers from which to assemble the desired sequence as well as primer variants per mutant.

1. Fluorescent primer: The primer should include a fluorophore [such as FAM (6-fluorescein amidite)] attached to the 5´ end that can be read by the capillary sequencer. We include a poly(A) stretch in the 5´ end of 20 nucleotides, which enables rapid and quantitative purification by poly(dT) beads. This stretch is followed by the reverse complement of the primer binding region of the RNA (see Figure 1 and Note 2 for the sequence of the fluorescent primer we use). This primer can be ordered from any oligo-synthesizing company, such as Integrated DNA Technologies (IDT).
2. Sequence to assemble: This sequence is generally composed of a T7 promoter in the 5´ end, followed by a buffer region starting with two guanines to allow for transcription, the sequence of the RNA of interest, a short 3´ buffer region, and the primer binding region (see Figure 1; also, see Note 2 for the primer binding region sequence we use). Check that adding these sequences do not potentially alter the structure of the RNA of interest by inspecting the minimum energy secondary structures of the RNA with and without the additional sequences as predicted by a secondary structure prediction program, such as RNAstructure (*25*). Lack of perturbation of the sequence of interest can be tested empirically by using an alternate set of primers and buffers and primer binding sites.
3. The NA_Thermo MATLAB toolkit (Das et al. unpublished software; freely available at https://simtk.org/home/na_thermo), or similar software (*26*, *27*) can be used to design the primers to assemble the sequence designed in the above step as well as the mutants of interest. The primers for the wild type sequence should be prepared or ordered first; all the steps below from PCR assembly to RNA transcription to sequencing ladders should be carried out to confirm that the primers produce the desired RNA. To obtain primer sequences using NA_Thermo:
    a. Download the toolkit from https://simtk.org/home/na_thermo and unpack in any directory. As with any MATLAB toolkit, the unpacked directory should be added to the MATLAB path.
    b. In MATLAB, type

        ```
        primers = design_primers(sequence)
        ```

        where *sequence* is your designed sequence from step 2. This will output (and save in the *primers* variable) the primers needed to build the designed sequence using

PCR assembly. The design_primers function also accepts parameters such as desired melting temperature between primers and primer size. To easily obtain plate layouts of the mutant library, first type:

```
sequences_to_order = single_mutant_library( primers, sequence, offset, region, libraries, name );
```

Where *offset* is an integer that is added to the sequence index to arrive at the conventional numbering for the RNA (-10 in the MedLoop example of Figure 1), *region* is the region of the sequence to be mutated (normally excluding the T7 promoter, buffer, and primer binding regions; 51:70 in the MedLoop example of Figure 1), *libraries* is a vector with integers that code for the type of libraries to generate, that is, to which base each nucleotide should be mutated (we usually mutate to the complement of each base, which corresponds to library 1; [1] in Figure 1), and *name* is an assigned name of the sequence. Finally, type:

```
output_sequences_to_order_96well_diagram( sequences_to_order, primers, name );
```

This will generate tab-delimited files with the sequences for each primer for each mutant, as well as images depicting the plate layouts (see Figure 1 for a plate layout example of a simple RNA and Figure 2 for a larger RNA with a more complex assembly). You should peruse the files in, e.g. Microsoft Excel, to confirm that the mutants are in the desired region.
  c. The primers obtained above can be prepared in 96-well format on standard synthesizers or ordered from any oligo synthesis companies, such as IDT.
4. The plate layouts generated in the step above are filled only in wells where the mutant primers differ from the wild type primers (see, e.g. green wells in Figures 1 and 2). All other wells should be filled with the appropriate wild type primer (for example, the white wells in Plate 1-Primer 3 and Plate 2-Primer 3 of Figure 2C should be filled with the wild type primer 3). See below for further details on PCR assembly of the primers.

**Mutant DNA assembly components**
1. 1200 µL of 5X HF (High Fidelity) buffer and 120 µL of 2000 units/mL Phusion polymerase per sample plate (Finnzymes, Thermo Scientific).

3. Standard reagents needed for agarose gel electrophoresis. We use Tris-Borate/EDTA buffer (TBE, Ambion) with 0.5 µg/mL ethidium bromide, and 96-well gel casting systems.

**RNA transcription**
1. 10X transcription buffer: 400 mM Tris-HCl, pH 8.1; 250 mM $MgCl_2$; 35 mM spermidine; 0.1% Triton X-100. Filter the buffer using a sterile 60 mL syringe through a 0.2 µm filter (Cole-Parmer). 10 mL of 10x transcription buffer can be used for at least thirty 96 well sample plates, and can be stored frozen at –20 °C for at least six months.

2. 120 µL of of 1 M DTT per sample plate.

3. 300 µL of 10 mM nTPs per sample plate.

4. 300 µL of 40% PEG 8000 per sample plate.

5. 30 µL of T7 RNA polymerase (New England Biolabs) per sample plate.

**RNA Folding**

1. 240 µL of 0.5 M Na-HEPES, pH 8.0, per sample plate. Stock should be prefiltered with a 0.2 µm filter. Other buffers can be used, but note that the acylation reaction rate decreases by 10-fold with each decreasing unit of pH.

2. 240 µL of 100 mM $MgCl_2$ per sample plate, if needed for proper folding of the nucleic acid system.

3. The reagents above may be different depending upon the conditions desired for proper folding/assembly of the RNA or DNA/RNA hybrid. For example, you can use a different buffer, different ions, and/or, additionally, small molecules that bind the RNA.

**Chemical mapping solutions**

Different chemical modifiers can be used to interrogate different aspects of a nucleic acid of interest. Each one affects different parts of the nucleic acid in different ways. The solutions needed to quench each reaction vary from modifier to modifier and are added after chemical modification has occurred. The following chemical modifiers are the ones most commonly used for nucleic acid structure probing; volumes prepared using these recipes can be used for up to two 96-well sample plates. Ideally, the molecules should be modified with single-hit frequency; the necessary reaction conditions will vary depending on temperature, time of the chemical modification step, and length of the RNA (see Note 5). The conditions should be optimized ahead of time on, e.g., the wild type sequence of the RNA. To avoid confusion between modification mixes or quenches, we recommend choosing only one of the following modifiers to carry out the mutate-and-map experiment and focusing on no more than two 96-well plates (192 constructs) during a given experimental session. Following are recipes for 4x concentrated modifier mixes, and concentrated quenches (see below for final amounts used in each reaction).

1. DMS mix for dimethyl sulfate mapping, which primarily gives a signal for exposed Watson-Crick edges of adenines and cytosines. Mix 10 µL fresh dimethyl sulfate (Sigma-Aldrich) into 90 µL 100% ethanol, then add 900 µL of sterile water. The quench for this reaction is 2-mercaptoethanol (Sigma-Aldrich).

2. CMCT mix for carbodiimide mapping using 1-cyclohexyl-(2-morpholinoethyl)carbodiimide metho-p-toluene sulfonate (CMCT; Sigma-Aldrich), which primarily gives a signal for exposed Watson-Crick faces of guanines and uracils in RNA and DNA. Mix 42 mg/ml of CMCT in $H_2O$. This reagent is usually kept at –20 °C. Before mixing, let the solid stock come to room temperature for 15 minutes before use. The quench for this reaction is 0.5 M Na-MES, pH 6.0.

3. NMIA mix for 2´-OH acylation mapping using N-methyl isatoic anhydride (NMIA; Sigma-Aldrich), which primarily gives a signal at dynamic RNA nucleotides: Mix 24 mg/mL in anhydrous dimethyl sulfoxide (DMSO). The quench for this reaction is 0.5 M Na-MES, pH 6.0. SHAPE reactions can also be quenched with 2-mercaptoethanol (*18*).

4. 1M7 mix for 2´-OH mapping using 1-methyl-7-nitroisatoic anhydride (1M7), a fast-acting reagent which primarily gives a signal at dynamic RNA nucleotides (*28*): Mix 8.5 mg/mL of 1M7 in anhydrous DMSO. The quench for this reaction is 0.5 M Na-MES, pH 6.0, or 2-mercaptoethanol.

5. It is important to make sure that enough quench mix to stop the chemical modification reactions is available: 1 mL of quench lasts for up to two 96-well sample plates.

**Chemical quench and bead solution**

1. Clean oligo-dT beads, here "Poly(A) purist" magnetic beads (Ambion/Applied Biosystems). Clean the beads (to remove any preservatives in the supplied stock solution) by taking out 200 µL of bead stock solution, adding 40 µL of 5 M NaCl, and separating them on a magnetic stand. Let the beads collect into a pellet and remove supernatant. Resuspend the beads in 200 µL sterile water, separate again, and remove the supernatant. Resuspend in 200 µL sterile water. This clean bead stock lasts for at least two weeks if kept at 4 °C.

2. Fluorescent primer: Prepare a 0.25 µM solution of a fluorescent primer (labeled at its 5´ end with fluorescein, available as a modification from Integrated DNA Technologies or other synthesis companies) that is complementary to the 3' end of the RNA of interest. In the following protocol, this primer should have a poly(A) stretch that will be used to bind to the oligo-dT beads in the purification steps immediately before and after chemical modification (Figure 3).

3. 8000 µL of 5 M NaCl per sample plate.

**Reverse transcription components**

1. 120 µL of 5X First Strand buffer per sample plate (Invitrogen).

2. 12 µL of 0.1 M DTT per sample plate.

3. 12 µL of Superscript III reverse transcriptase (Invitrogen) per sample plate.

**RNA hydrolysis and acid quench solutions**

1. 840 µL of 0.4 M NaOH per sample plate. This 0.4 M NaOH can be prepared in 50 mL volumes and kept at room temperature for at least two months.

2. Acid quench mix, prepared by mixing 1 volume of 5 M NaCl, 1 volume of 2 M HCl, and 1.5 volumes of 3 M sodium acetate (pH 5.2). This stock can be kept at room temperature for at least two months. See Note 10.

**ROX-Formamide elution**

1. ROX-Formamide mix: Mix 2.75 µL of ROX 350 ladder in 1200 µL of Hi-Di formamide (both available from Applied Biosystems). This volume is enough for one 96-well sample plate; larger stocks can be prepared and stored frozen at –20°C.

## *Methods*

All procedures are performed at room temperature unless specified. For pipetting mixes into each well of a 96 well plate, use an extra plate and a multi-channel pipetter to reduce pipetting efforts: pre-aliquot the mix to 8 wells in a separate, auxiliary 96 well plate (see Notes 3 and 4, and additional reminders below). Then use an 8-channel pipetter to pipette the desired amount per well to the 12 columns in the main reaction plate.

## Mutant DNA assembly

Primer assembly by PCR is used to produce the DNA to transcribe RNA for all mutants. This strategy allows for rapid generation of single-nucleotide mutants – to produce a mutant for a specific position, only a few of the primers in the assembly need to be changed – and is amenable to the 96-well plate format. To facilitate mixing the correct primers together, it is recommended to organize the primers for all mutants into 96-well plates: each plate should contain all variants of a given primer in the assembly for all mutants with wild type primers filling the other well (see the *Sequence section assembly primers* in *Materials* above and Figures 1 and 2),

1. Dilute the primer stocks to the appropriate concentration. Usually, the first and last primers in the assembly are set to 100 µM concentration, and any intermediate primers are set to a lower initial concentration of (e.g. 1 µM). The final concentration of these primers in the PCR reactions will be 25-fold less.

2. Pipette 2 µL of the first primer plate into a 96-well PCR plate using a multichannel pipette. With the addition of the rest of the PCR mix, this will give a final concentration of 4 µM. Repeat this procedure for the rest of the primer plates so that the first and last primers are at a concentration of 4 µM and the intermediate primers are at a final concentration of 40 nM. Here and throughout, dispose tips after each pipetting step to avoid cross-contamination.

3. Prepare PCR mix. Each 50 µL reaction will involve 10 µL of 5X HF buffer, 1 µL 10 mM dNTPs, 1 µL 2000 units/mL Phusion polymerase, and $38 - 2Y$ µL of sterile water, where $Y$ is the number of primers in the assembly, per reaction (e.g. for an assembly of 2 primers, 34 µL of water would be added for each reaction). Scale the volumes by 1.2 times the number of samples to allow for volume lost while pipetting.

4. Add $50 - 2Y$ µL of the PCR mix into each well of the PCR plate, where $Y$ is the number of primers in the assembly, to give 50 µL reactions. It is easiest to use the "auxiliary" plate with a column filled with the PCR mix and pipette to the whole plate from there.

5. Cover the plate with a PCR plate seal and put it on a thermocycler with the following program:

    | Steps | Time/Cycles | Temperature |
    |---|---|---|
    | Denaturation | 30 seconds | 98°C |
    | Denaturation | 10 seconds | 98°C |
    | Annealing | 30 seconds | 64°C |
    | Extension | 30 seconds | 72°C |
    | Repeat step 2-4 | 29 cycles more | |
    | Polymerization | 10 minutes | 72°C |

    The annealing temperature may need to be adjusted based on the primer melting temperatures under the PCR conditions; the above table uses Finnzyme/Thermo instructions for Phusion (+4°C over the melting temperature of primers, which are set to be greater than or equal to 60°C in NA_thermo primer design scripts).

6. Assembly of the DNA templates with the correct length may be confirmed by agarose gel electrophoresis of 10 μL aliquots and use of a 20 bp length standard (e.g., from Fermentas). If a 96-sample gel-casting system is not available, a subset of mutants can be chosen for agarose gel electrophoresis.

## DNA purification

The purification method described here yields approximately 2 μg linear DNA template per sample well.

1. After the PCR is completed, add 72 μL (1.8 volumes) Agencourt AMPure XP beads to 40 μL of the PCR mix in each well, mix by pipetting up and down and leave at room temperature for 10 minutes. For purifying constructs smaller than 100 nucleotides, see Note 6.

2. Separate the beads by setting the plate on a 24 post-magnetic stand for 7 minutes. Discard supernatant and rinse each sample with 200 μl of 70% ethanol (see Note 1); incubate at room temperature for 1 minute and discard.

3. Repeat the previous 70% ethanol wash, discard the supernatant 70% ethanol and leave the beads to dry for at least 15 minutes on the magnetic stand. Residual ethanol will impair transcription.

4. Take the plate off the magnetic stand and elute each sample in 35 μl of sterile water. After 5 minutes, re-separate the beads using the magnetic stand.

5. Pipette out the solution to the new plate, leaving the magnetic beads behind.

6. The concentration of DNA can be checked via UV absorbance at 260 nm on, e.g., a Nanodrop system.

## RNA transcription and purification

The following transcription and purification method results in approximately 300 pmoles of RNA per sample well.

1. Prepare DNA dilutions by adjusting the stock concentration to be 0.2 μM. Aliquot 2.5 μl of the DNA dilutions into a new 96-well plate.

2. Prepare the transcription mix. Each 25 μL transcription will involve 2.5 μL of 10X transcription buffer, 1 μL 1M DTT, 2.5 μL 10 mM NTPs, 2.5 μL 40% PEG 8000, 13.75 μL sterile water, and 0.25 μL T7 RNA polymerase per sample. Scale the volumes of these shared components by 1.2 times the number of samples to account for pipetting errors (e.g. around 2880 μL for 96 samples).

3. Add 22.5 μL of transcription mix into each well and mix well by pipetting up and down.

4. Incubate at 37 °C for 3 hours – for best results, use a water bath or a thermocycler for the incubation.

5. Successful RNA transcription can be confirmed by denaturing agarose gel electrophoresis. (e.g. formaldehyde and agarose gel), although RNA bands may not be clear due to incomplete denaturing. Alternatively, reverse transcription provides a check on the RNA quality and sequence, but occurs downstream (see below).

6. Purify the RNA following the same steps as in DNA purification (for purification of small constructs, see Note 6). Residual low amounts of DNA template does not interfere in later steps. However, if desired, DNAse I and calcium-containing DNAse I buffer can be added to the transcription mix before purification to degrade the DNA template.

7. Measure the concentration of each RNA mutant via UV absorption on, e.g., a Nanodrop.

**Folding or complex assembly**

1. Prepare folding buffer: Mix 2 µL of 0.5 M Na-HEPES pH 8, 2 µL 100 mM $MgCl_2$, and any additional reagents needed for proper folding of the RNA in question, per sample.

2. Transfer $4 + X$ µL of folding buffer to each well in a new 96-well plate per sample plate, where $X$ is the volume of additional reagents used in the folding buffer – add $Y$ µL of nuclease free water such that $4 + X + Y = 14$ µL. Use the 96-well "auxiliary" plate to transfer the corresponding volume to the sample plate using a multichannel pipetter.

3. Prepare a plate of diluted RNA stocks at a concentration of at least 0.12 µM. Transfer 1 µL of each RNA mutant to the folding buffer plate to achieve at least 12 picomoles of RNA molecules. The total volume in each well will be 15 µL.

4. Equilibrate at room temperature for 20 minutes or incubate the plate at the temperature and conditions needed to fold the RNA.

**Chemical modification**

1. Add 5 µL of the freshly prepared chemical modification solution to each sample well and mix well by pipetting up and down. It is again easiest to pre-aliquot this buffer to the separate "auxiliary" plate (e.g., 65 µL in 8 wells), and then using the multichannel pipetter to mix into the reaction plate.

2. Cover the sample plate with sealer tape and incubate at room temperature. The incubation time with the modification reagent should be varied according to the length of the RNA and the chemical modifier used (e.g. for a 200 nucleotide RNA at 24 °C, DMS and CMCT requires an incubation time of 15 minutes; 1M7 modification requires 5 minutes, while NMIA requires 30 minutes). Modifier concentration or modification time should be decreased for longer RNAs to maintain single-hit kinetics on a longer construct (Note 5).

**Quench and purification**

1. Prepare bead-quench mix: Mix 3 µL of 5 M NaCl, 1.5 µL of clean oligo-dT beads, 5 µL of quench reagent (see above) and 0.25 µL of the 0.25 µM fluorescent primer stock (9.75 µl final volume) per sample. Scale the volumes according to 1.2 times the number of samples (e.g. around 1120 µL for 96 samples). See Note 7 for reactions using a 2-mercaptoethanol quench; and Note 8 for NMIA reactions.

2. Remove and discard the seal from the sample plates. Add 9.75 µL of the bead-quench mix to all sample wells (see Figure 4A). As before, pre-aliquot the quench mix to a set of 8 wells in the separate "auxiliary" plate and use an 8-channel pipetter to transfer the mix. Wait 7 minutes to allow the primer to bind to the RNA.

3. Set the sample plate on a 96-post magnetic stand and wait 7 to 10 minutes for the beads to collect (Figure 4B).

4. Remove and discard supernatant while the sample plate is still on the separator. After discarding supernatant, rinse the beads by applying 100 µL of 70 % ethanol onto each well (see Note 1). The beads will stay collected at the bottom of the plate. Wait 1 minute and remove the ethanol from samples while the sample plate is on magnetic separator.

5. Repeat the 70% ethanol wash above: add 100 µL, wait 1 minute, and remove while keeping plate on the magnetic separator. In this second wash it is important to remove the 70% ethanol supernatant thoroughly to avoid affecting the reverse transcription step. This can be accomplished by removing most of the ethanol with a P100 multichannel pipetter and subsequently removing any remaining droplets with a multichannel P10 pipetter placed close to the beads collected at the bottom of the plate.

6. Let the sample plate dry for 10 minutes while still on the magnetic separator. After the samples are dry, make sure to re-inspect the wells for any residual ethanol. If any drops are visible, pipette them out and leave to dry for 5 minutes longer. During this time, prepare reverse transcription mix (see next section).

7. Resuspend each sample in 2.5 µL of RNase free water and take the plate off the magnetic stand. If reverse transcription mix is not ready, adhere a fresh plastic seal to plate to prevent evaporation.

### Reverse transcription and final purification

1. Prepare the Superscript III reverse transcription mix: Mix 1 µL of 5X First Strand buffer, 0.25 µL of 0.1 M DTT, 0.4 µL of 10 mM dNTPs, 0.75 µL sterile water, and 0.1 µL Superscript III reverse transcriptase per sample (the final volume of mix per sample will be 2.5 µL ). Scale the volumes accordingly for approximately 1.2 of the number of samples (e.g. around 300 µL for 96 samples). See Note 9 for reaction mixture to create sequencing ladders.

2. Take the sample plate off the magnetic separator, and add 2.5 µL of the Superscript III mix to the sample wells. The total volume will be 5 µL. Again, pre-aliquot larger amounts to eight wells in the separate "auxiliary plate", and use the 8-channel pipetter to transfer the mix. Mix very well at least 10 times, or until the beads go fully back into solution. Make sure the reaction plate is not near the magnets for this step.

3. Seal the sample plate, and put it in an incubator set to 42°C for 30 minutes. For longer RNAs, leaving the samples for longer times may lead to slightly better reverse transcriptase extension.

4. To remove the RNA template after reverse transcription, add 5 µL of 0.4 M NaOH to each well (again use the "auxiliary" plate to reduce pipetting effort). Incubate the samples at 90°C for three minutes to hydrolyze RNA, while leaving fluorescent cDNA behind.

5. Take plate out of the incubator and cool on ice for 3 minutes.

6. Add 5 µL of acid quench mix (see Note 10) to each well, using the "auxiliary" plate to reduce pipetting effort.

7. After 1 minute, put the sample plate on a magnetic stand (either 24 or 96 post formats are fine) and wait 7 minutes until the beads aggregate. Remove and discard supernatant with a multichannel pipetter.

8. Residual salt in the wells may interfere when injecting the samples into the capillary sequencer. To remove these impurities, rinse the beads by applying 100 µL of 70% ethanol onto each well. Wait one minute and remove the ethanol.

9. Repeat the 70% ethanol wash above. Remove the residual ethanol thoroughly and let the plate dry for 10 minutes.

10. Resuspend the beads in 11 µL of ROX-Formamide mix while the plate is still in the magnetic stand. Wait 15 minutes to complete the elution.

11. Transfer the supernatant to a capillary sequencing optical plate and remove any bubbles by centrifuging the plate with a benchtop plate centrifuge.

12. The samples are now ready to run in an ABI 3100 capillary sequencer or similar with a standard sequencing or fragment analysis protocol.

### Data analysis, curation, and publication

The resulting data after the capillary electrophoresis run, saved in one ab1 file per sample well, can be analyzed using HiTRACE (*29*), ShapeFinder (*30*), CAFA (*31*), or other similar software packages. Hereafter we assume that the HiTRACE MATLAB toolkit is used to analyze the data, as it has been optimized for quantitation of data sets involving hundreds of traces that benefit from alignment and global sequence annotation.

1. HiTRACE is a MATLAB toolkit that helps align, normalize, map to sequence, and quantify the data obtained from chemical mapping experiments of RNA, including mutate-and-map assays. The full procedure of analyzing chemical mapping data is given in the HiTRACE documentation and tutorials – see, e.g., https://sites.google.com/site/rmdbwiki/hitrace. The manual covers everything from installing the toolkit to analyzing the raw capillary data, to saving the quantified chemical reactivities in the RDAT annotated file format. Once an RDAT file with the experimental data is obtained, tools from the RNA Mapping Database (RMDB) (*32*) can be used to share, visualize, and obtained computationally predicted secondary structure models of the RNA guided by the mutate-and-map data.
2. Researchers are encouraged to share their analyzed data sets by submitting them to the RMDB or SNRNASM (*33*) repositories. To submit to the RMDB, go to http://rmdb.stanford.edu/repository/register/ to register to the site and to http://rmdb.stanford.edu/repository/submit/ to submit your data. After submission and approval by the RMDB curators, the entry can be visualized in its details page (see Figure 5A).
3. The mutate-and-map data can be used to guide secondary structure algorithms, such as RNAstructure (*25*). Calculating the Z-scores across each nucleotide positions reveals the most significant perturbations due to each mutation. This information can be used as bonuses in the secondary structure prediction algorithms to yield improved secondary structure models (*24*). Further, repeating this procedure by resampling the data with replacement using bootstrapping gives helix-wise confidence estimated (*19, 24*) . This pipeline is available at the RMDB structure server; to obtain mutate-and-map guided secondary structure models:

   a. Visit http://rmdb.stanford.edu/structureserver/ . In "Input options" click "Upload RDAT" to expand the RDAT uploading options. In the expanded pane, click choose file and choose the RDAT file containing the mutate-and-map data. When the file is uploaded, the fields in the structure server forms will be auto-populated with the annotations and data of the uploaded RDAT file.
   b. In "Bonus options", click "2D bonuses" to expand the 2D bonus option pane. You will see the data as a matrix of bonuses in "Input bonuses" (some entries in this matrix will be zeros if some nucleotides in the sequence are missing data). Check "Filter by Z-sores" and input the number of bootstrap iterations for obtaining helix-wise confidence values. The larger the iterations, the longer it will take but the more accurate the confidence estimates will be. For most RNAs. 100 to 400 iterations appears sufficient to get converged bootstrap confidence values.
   c. In "Other options" adjust the temperature if necessary and inspect the (optional) reference secondary structure, given in dot-parentheses notation. Click on "Submit"; you will be directed to a page containing the predicted model, including confidence estimates (see Figure 5B).

## *Notes*

1. It is easiest to have a large amount of 70% ethanol stored in a sterile 96-well rack in which each well holds at least 2 mL of volume. Care should be taken to avoid ethanol evaporation, as lower ethanol content washes will remove bound cDNA sample. Seal the reservoir when not in use during the protocol, discarding and freshly filling its contents before the experiment starts and before each major wash step in the protocol.

2. In the example experiments above we use AAAAAAAAAAAAAAAAAAAAGTTGTTGTTGTTGTTTCTTT as our fluorescent primer sequence. Furthermore, we add a T7 promoter (TTCTAATACGACTCACTATA) and a buffer of "unstructured" nucleotides beginning with two guanines (e.g. GGCCAAAACAACGGAA) at the 5' end of our sequence of interest to make the RNA amenable to *in vitro* transcription. Similarly, we add a buffer, unstructured region of approximately 10 nucleotides and a "tail" (AAAACAAAACAAAGAAACAACAACAACAAC) that serves as the binding region for the fluorescent primer. See examples sequences for these buffers in Figure 1 and 2.

3. The 96-well format of the protocol is best exploited when using an "auxiliary" plate to pre-aliquot premixes. This allows the use of multichannel pipetters to transfer the mixes to the sample plates with less pipetting effort.

4. It is best practice not to reuse pipette tips when using the multichannel pipetters in any step of the protocol other than the ethanol washes to avoid sample cross-contamination.

5. To maintain single hit kinetics on large RNA constructs, concentration of the chemical modifier should be reduced. We suggest that 1/2 the concentration of the chemical modifier be used for RNAs with lengths between 100 to 150 nucleotides, and 1/4 for RNAs with lengths between 150 and 250.

6. Ampure XP purification of small nucleic acids can be optimized by adding polyethylene glycol (PEG) 8000 to the Ampure beads. We have found that a mix of 3 volumes 40% PEG 8000 with 7 volumes of Ampure beads optimal for recovering DNA and RNA of 50 nucleotides.

7. The 2-mercaptoethanol quenching reagent can interfere with magnetic bead aggregation if left for too long before performing the first 70% ethanol wash after chemical modification. It is thus recommended to mix the quench premix right before applying it. Ruined samples will appear viscous and green. Samples washed timely and properly will appear light brown.

8. For NMIA probing, after performing chemical modification, the beads may smear against the well walls and will refuse to collect at the bottom of the well when the plate is transferred to the 96-post magnet. It is best not scrape the beads off the walls. Instead, pipette up and down against the walls to wash them into the bottom of the well.

9. As a quality check and also to prepare references for assigning bands in the capillary electrophoresis traces, acquisition of Sanger sequencing ladders for at least the wild type construct are recommended. Prepare 1.2 pmols RNA, 1.5 µL oligo(dT) beads, 0.25 µL 0.25 µM fluorophore-labeled primer, and water up to 2.5 µL volume for each sequencing reaction. Then add the following 2.5 µL mix: 1.0 µL of 5X First Strand buffer, 0.25 µL of 0.1 M DTT, 0.4 µL of 1.0 mM dNTPs, 0.35 µL sterile water, 0.4 µL of 1 mM ddATP (or one of the other three 2´-3´-dideoxynucleotide triphosphates) and 0.1 µL Superscript III reverse transcriptase per sample (the final volume will be 5 µL). Carry out remaining reverse transcription and purification steps as described in "Chemical Mapping".

10. It is critical that the acid quench neutralizes the 0.4 M NaOH before the last wash. It is recommended to confirm that the final pH of a mix of 5 µL of 0.4 M NaOH and 5 µL of acid quench mix is between 5 and 7.

**Acknowledgments**

We thank A. Becka, T. Mann, and S. Tian for comments on the manuscript. This work is supported by the Burroughs-Wellcome Foundation (CASI to R.D.), a CONACyT fellowship (to P.C.), and the National Institutes of Health (T32 HG000044 to C.C.V. and R01 GM102519 to R.D.).

## *References*

1. Mandal, M., Lee, M., Barrick, J. E., Weinberg, Z., Emilsson, G. M., Ruzzo, W. L., and Breaker, R. R. (2004) A glycine-dependent riboswitch that uses cooperative binding to control gene expression., *Science (New York, N.Y.) 306*, 275–9.

2. Kulshina, N., Baird, N. J., and Ferré-D'Amaré, A. R. (2009) Recognition of the bacterial second messenger cyclic diguanylate by its cognate riboswitch., *Nature structural & molecular biology 16*, 1212–7.

3. Gesteland, R. F., Cech, T. R., and Atkins, J. F. (2006) The RNA World 3rd ed. CSHL Press.

4. Leontis, N. B., and Westhof, E. (1998) The 5S rRNA loop E: Chemical probing and phylogenetic data versus crystal structure, *RNA 4*, 1134–1153.

5.  Staley, J. P., and Guthrie, C. (1998) Mechanical Devices of the Spliceosome: Motors, Clocks, Springs, and Things, *Cell 92*, 315–326.

6.  Lukavsky, P. J., Kim, I., Otto, G. A., and Puglisi, J. D. (2003) Structure of HCV IRES domain II determined by NMR, *Nature Structural Biology 10*, 1033–1038.

7.  Brostoff, S. W., and Ingram, V. M. (1967) Chemical modification of yeast alanine-tRNA with a radioactive carbodiimide., *Science 158*, 666–669.

8.  Moazed, D., Stern, S., and Noller, H. F. (1986) Rapid chemical probing of conformation in 16 S ribosomal RNA and 30 S ribosomal subunits using primer extension, *Journal of Molecular Biology 187*, 399–416.

9.  Wilkinson, K. A., Merino, E. J., and Weeks, K. M. (2006) Selective 2'-hydroxyl acylation analyzed by primer extension (SHAPE): quantitative RNA structure analysis at single nucleotide resolution., *Nature protocols 1*, 1610–6.

10. Lucks, J. B., Mortimer, S. A., Trapnell, C., Luo, S., Aviran, S., Schroth, G. P., Pachter, L., Doudna, J. A., and Arkin, A. P. (2011) Multiplexed RNA structure characterization with selective 2'-hydroxyl acylation analyzed by primer extension sequencing (SHAPE-Seq)., *Proceedings of the National Academy of Sciences of the United States of America* (Hage, J., and Meeus, M., Eds.) *108*, 11063–11068.

11. Sood, V. D., Beattie, T. L., and Collins, R. A. (1998) Identification of phosphate groups involved in metal binding and tertiary interactions in the core of the Neurospora VS ribozyme., *Journal of molecular biology 282*, 741–50.

12. Rhee, Y., Valentine, M. R., and Termini, J. (1995) Oxidative base damage in RNA detected by reverse transcriptase, *Nucleic Acids Research 23*, 3275–3282.

13. Chow, C. S., Cunningham, P. R., Lee, K., Meroueh, M., SantaLucia, J., and Varma, S. (2002) Photoinduced cleavage by a rhodium complex at G·U mismatches and exposed guanines in large and small RNAs, *Biochimie 84*, 859–868.

14. Kertesz, M., Wan, Y., Mazor, E., Rinn, J. L., Nutter, R. C., Chang, H. Y., and Segal, E. (2010) Genome-wide measurement of RNA secondary structure in yeast., *Nature 467*, 103–7.

15. Underwood, J. G., Uzilov, A. V, Katzman, S., Onodera, C. S., Mainzer, J. E., Mathews, D. H., Lowe, T. M., Salama, S. R., and Haussler, D. (2010) FragSeq: transcriptome-wide RNA structure probing using high-throughput sequencing., *Nature methods 7*, 995–1001.

16. Wells, S. E., Hughes, J. M., Igel, A. H., and Ares, M. (2000) Use of dimethyl sulfate to probe RNA structure in vivo., *Methods in Enzymology 318*, 479–493.

17. Tijerina, P., Mohr, S., and Russell, R. (2007) DMS footprinting of structured RNAs and RNA-protein complexes., *Nature Protocols 2*, 2608–2623.

18. Spitale, R. C., Crisalli, P., Flynn, R. A., Torre, E. A., Kool, E. T., and Chang, H. Y. (2012) RNA SHAPE analysis in living cells., *Nature chemical biology 9*, 18–20.

19. Kladwang, W., VanLang, C. C., Cordero, P., and Das, R. (2011) Understanding the errors of SHAPE-directed RNA structure modeling., *Biochemistry 50*, 8049–56.

20. Cordero, P., Kladwang, W., VanLang, C. C., and Das, R. (2012) Quantitative dimethyl sulfate mapping for automated RNA secondary structure inference., *Biochemistry 51*, 7037–9.

21. Washietl, S., Hofacker, I. L., Stadler, P. F., and Kellis, M. (2012) RNA folding with soft constraints: reconciliation of probing data and thermodynamic secondary structure prediction., *Nucleic Acids Research* 1–12.

22. Kladwang, W., and Das, R. (2010) A Mutate-and-Map Strategy for Inferring Base Pairs in Structured Nucleic Acids: Proof of Concept on a DNA/RNA Helix, *Biochemistry 49*, 7414–7416.

23. Kladwang, W., Cordero, P., and Das, R. (2011) A mutate-and-map strategy accurately infers the base pairs of a 35-nucleotide model RNA, *RNA 17*, 522–534.

24. Kladwang, W., VanLang, C. C., Cordero, P., and Das, R. (2011) A two-dimensional mutate-and-map strategy for non-coding RNA structure., *Nature chemistry 3*, 954–62.

25. Mathews, D. H., Disney, M. D., Childs, J. L., Schroeder, S. J., Zuker, M., and Turner, D. H. (2004) Incorporating chemical modification constraints into a dynamic programming algorithm for prediction of RNA secondary structure., *Proceedings of the National Academy of Sciences of the United States of America 101*, 7287–92.

26. Rydzanicz, R., Zhao, X. S., and Johnson, P. E. (2005) Assembly PCR oligo maker: a tool for designing oligodeoxynucleotides for constructing long DNA molecules for RNA production., *Nucleic acids research 33*, W521–5.

27. Thachuk, C., and Condon, A. (2007) On the Design of Oligos for Gene Synthesis, in *2007 IEEE 7th International Symposium on BioInformatics and BioEngineering*, pp 123–130. IEEE.

28. Mortimer, S. A., and Weeks, K. M. (2007) A fast-acting reagent for accurate analysis of RNA secondary and tertiary structure by SHAPE chemistry., *Journal of the American Chemical Society 129*, 4144–5.

29. Yoon, S., Kim, J., Hum, J., Kim, H., Park, S., Kladwang, W., and Das, R. (2011) HiTRACE: high-throughput robust analysis for capillary electrophoresis, *Bioinformatics 27*, 1798–1805.

30. Vasa, S. M., Guex, N., Wilkinson, K. A., Weeks, K. M., and Giddings, M. C. (2008) ShapeFinder: a software system for high-throughput quantitative analysis of nucleic acid reactivity information resolved by capillary electrophoresis., *RNA (New York, N.Y.) 14*, 1979–90.

31. Mitra, S., Shcherbakova, I. V, Altman, R. B., Brenowitz, M., and Laederach, A. (2008) High-throughput single-nucleotide structural mapping by capillary automated footprinting analysis., *Nucleic acids research 36*, e63.

32. Cordero, P., Lucks, J. B., and Das, R. (2012) An RNA Mapping Database for curating RNA structure mapping experiments., *Bioinformatics (Oxford, England) 28*, 3006–3008.

33. Rocca-Serra, P., Bellaousov, S., Birmingham, A., Chen, C., Cordero, P., Das, R., Davis-Neulander, L., Duncan, C. D. S., Halvorsen, M., Knight, R., Leontis, N. B., Mathews, D. H., Ritz, J., Stombaugh, J., Weeks, K. M., Zirbel, C. L., and Laederach, A. (2011) Sharing and archiving nucleic acid structure mapping data., *RNA (New York, N.Y.) 17*, 1204–12.

# Figures

**Figure 1:** Steps for designing the sequences necessary for building and probing an RNA with the mutate-and-map strategy. Here, the steps to design the assembly of the MedLoop RNA are shown. Step 0: Before designing the sequence that will be transcribed *in vitro*, a fluorescent primer for reverse transcribing the RNA is needed. This primer must have a fluorophore (such as FAM (6-fluorescein amidite)) attached to the 5´ end followed by a stretch of 20 adenines and ended by a sequence that is the reverse complement to the priming region of the RNA. The primer will bind to the poly(A) purist beads on the 5´ end to allow for easy purification and will be attached to the RNA on the 3´ end for reverse transcription. Step 1: Designing the sequence includes adding a T7 promoter, a 5´ buffer (normally beginning with two guanines to allow transcription of the RNA), the sequence of interest, a short 3´ buffer, and a sequence where the fluorescent primer will bind. Step 2: Primers required for the PCR assembly of the complete sequence can be obtained from software packages such as NA_thermo. Step 3: Example DNA primer plate layout for a library of single-nucleotide mutants and wild-type of a two primer assembly of the MedLoop RNA. Green wells mark where each primer varies due to the single nucleotide mutation while white wells use the respective wild type primers; grey wells are not used for any mutant and are left empty. Combining all primer plates per well into one plate using the appropriate concentrations and performing PCR will assemble all mutants simultaneously. Arrows to the first 8 wells (A1 to H1) indicate the mutant contained in that well (e.g. A01U, the mutant whose first base, an adenine, is mutated to a uracil is contained in well A2).

## Step 0: Design Primer

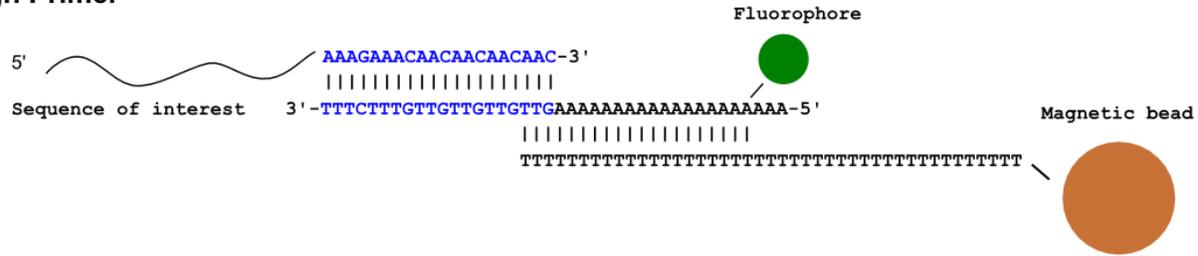

## Step 1: Design Sequence

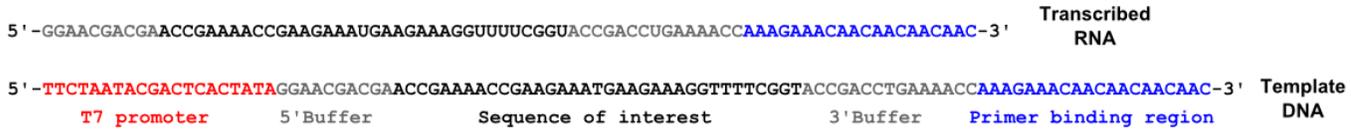

## Step 2: Design Assembly

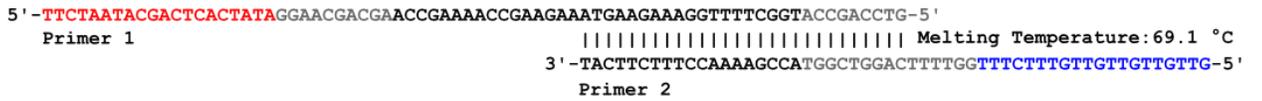

## Step 3: Design Mutants in 96-well plate format

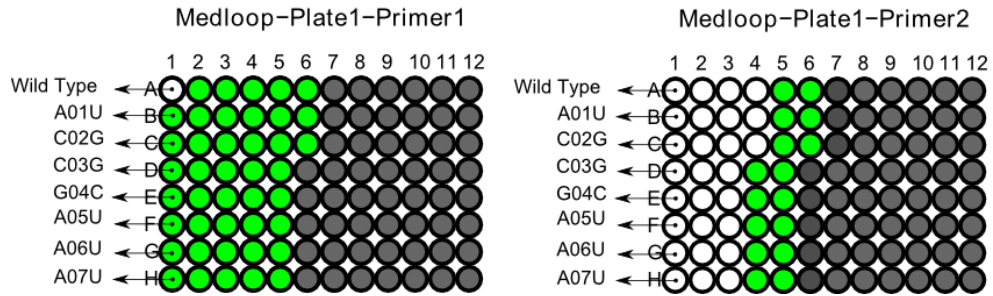

**Figure 2:** Mutant libraries for larger RNAs, such as the P4-P6 domain of the *Tetrahymena* group I intron ribozyme shown here, require several plates. In this case, the RNA of interest is assembled through 6 primers and requires almost two plates. (A) The sequence, with T7 promoter, buffer regions, and primer binding region included. (B) The six-primer assembly of the RNA. (C) Plate layouts for assembling the mutant library are shown with the same color coding as in Figure 1. The plate layouts for primers 5 and 6 for plate 1 and primers 1 and 2 for plate 2 are omitted since only wild-type primers are used for all mutants in those cases.

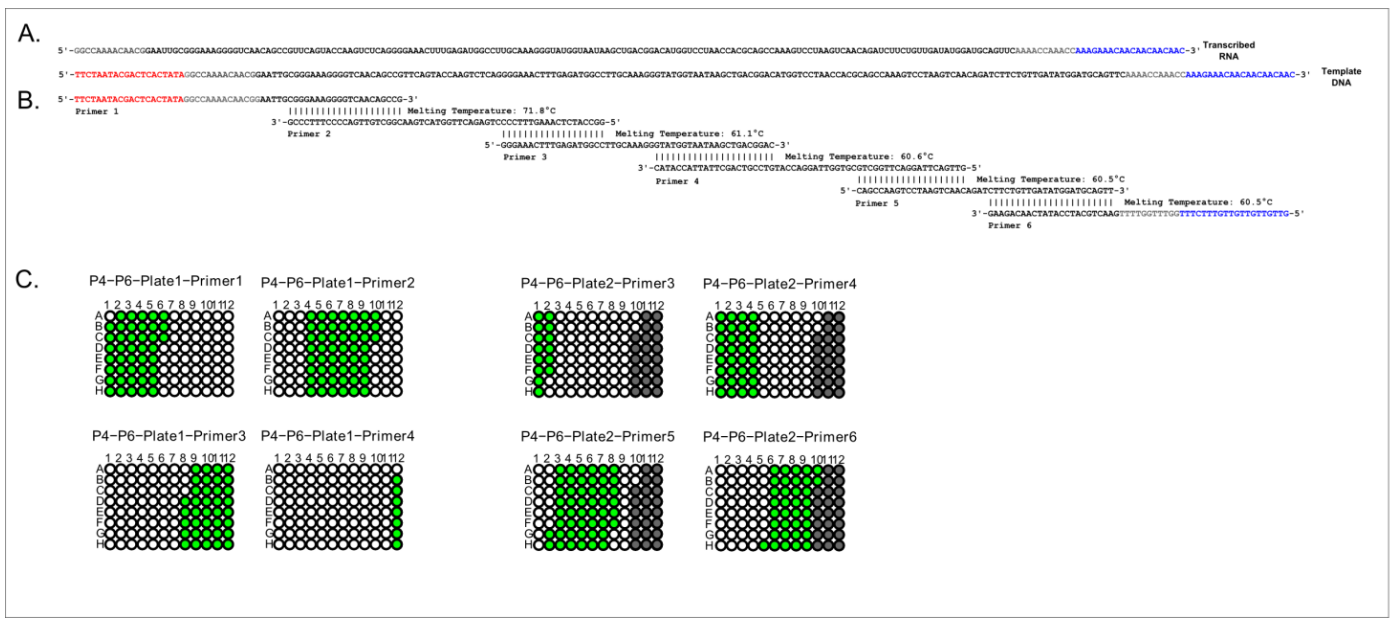

**Figure 3:** Workflow of the chemical modification for capillary electrophoresis using bead purification. The RNA is chemically modified and pulled down with oligo dT beads that bind the poly(A) tail of a fluorescent primer that contains a region complementary to the 3' of the RNA. The RNA is then reverse transcribed – the reverse transcriptase stops at modified sites and produces a modification pattern that can be mapped to the sequence of interest using capillary electrophoresis. The read-out can be simultaneously obtained for all mutants thanks to the 96-well format of the protocol. The resulting data is then analyzed using freely available software packages. Finally, the RNA Mapping Database can be used to share and visualize the data as well as to predict experimentally informed computational secondary structure models.

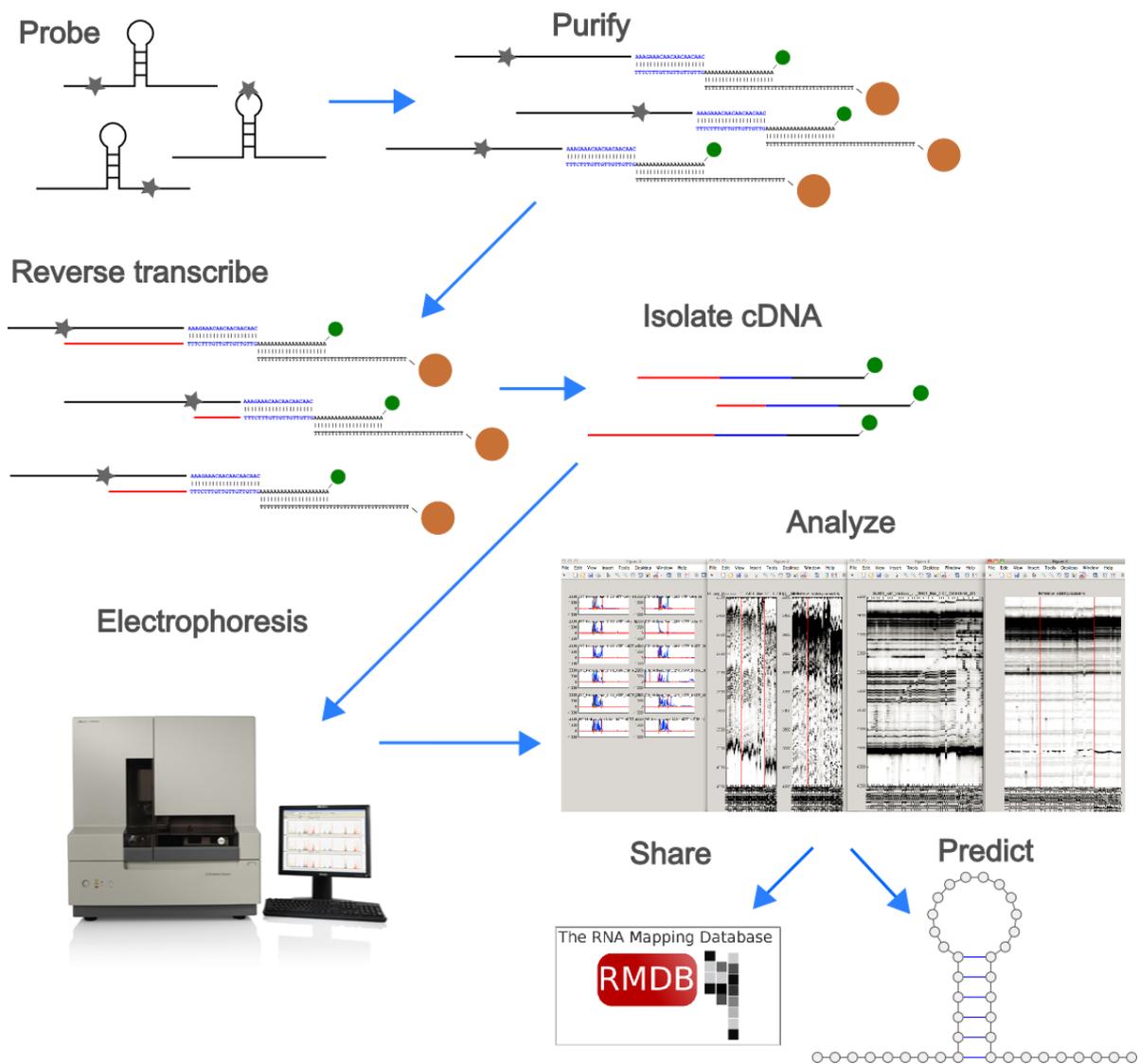

**Figure 4:** (A) Example wells after adding the bead-quench mix into modified RNA. (B) The sample beads collect into pellets after putting the sample plate on the magnetic stand for 7 min. Mutate-and-map experiments can include samples in all 96 wells of a plate; snapshots from a trial experiment with 16 wells are shown here.

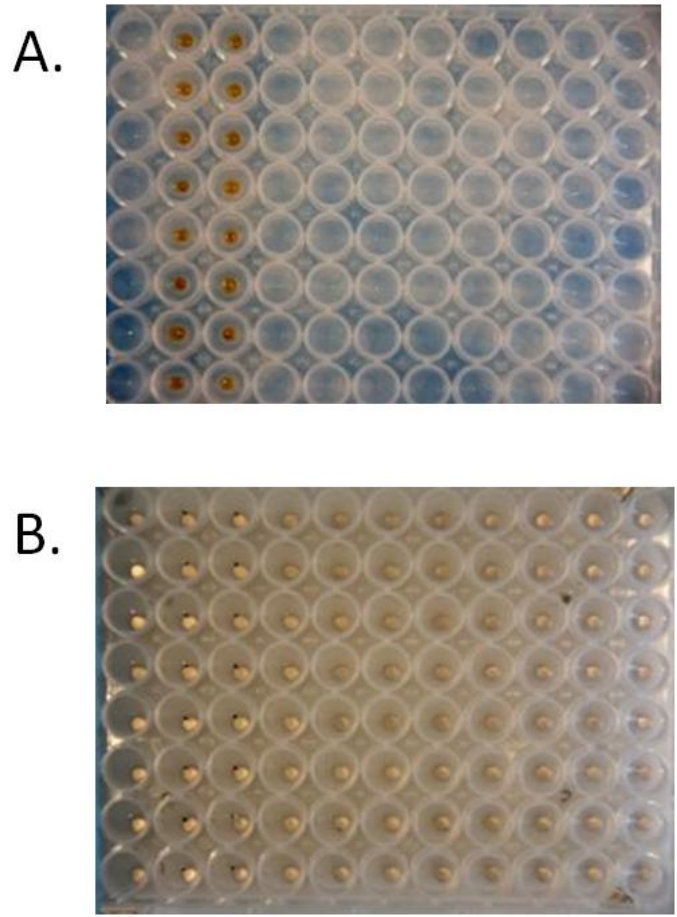

**Figure 5:** After data collection, reactivities of each position for each mutant can be quantified using Hi-TRACE. These data can be made publicly available through the RNA Mapping DataBase (RMDB). The RMDB provides tools to visualize, share, and use the data in producing secondary structure models of the RNA. Here, we show the (A) details page and (B) resulting model of the RMDB structure prediction server of the MedLoop motif (entry MDLOOP_SHP_0002) as well as the (C) details page and (D) resulting model of the P4-P6 domain of the *Tetrahymena* ribozyme (entry TRP4P6_SHP_0003).

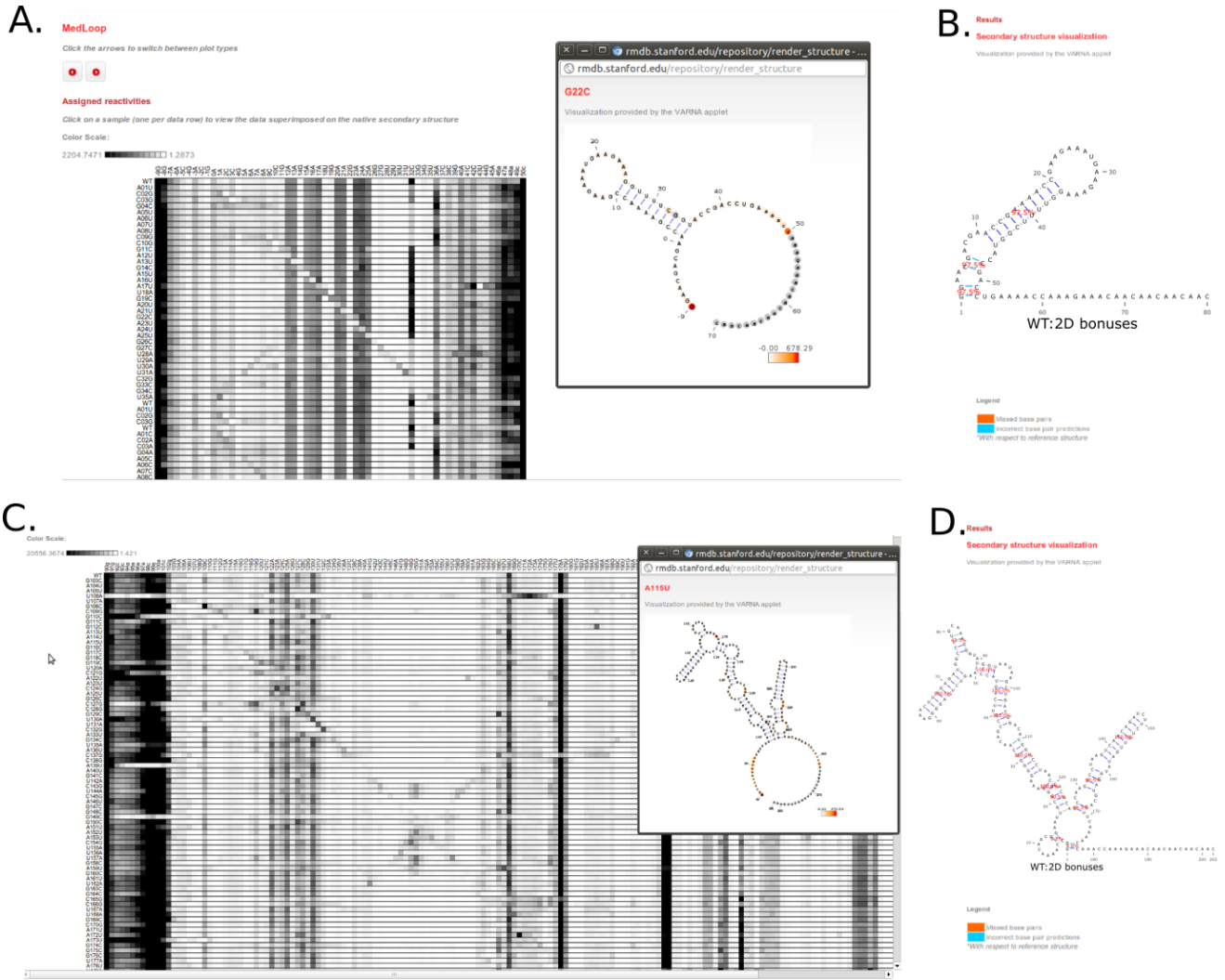